\newcommand{\CLASSINPUTtoptextmargin}{2cm}
\newcommand{\CLASSINPUTbottomtextmargin}{2cm}
\documentclass[conference,a4paper]{IEEEtran}
\ifCLASSINFOpdf
  \usepackage[pdftex]{graphicx}
\else
\fi

\usepackage{nicefrac}
\usepackage{graphicx}

\begin{document}
\title{Energy Savings When Migrating Workloads \\to the Cloud}

\author{\IEEEauthorblockN{Yan Zheng}
    \IEEEauthorblockA{Department of Electrical and Computer Engineering\\
        University of Delaware, Newark, DE 19711\\
        Email: yzheng@udel.edu
        \\and\\
        Cloudamize\\
        Philadelphia, PA 19103\\
        yzheng@cloudamize.com}
\and
\IEEEauthorblockN{Stephan Bohacek}
\IEEEauthorblockA{Department of Electrical and Computer Engineering\\
University of Delaware, Newark, DE 19711\\
Email: bohacek@udel.edu
\\and\\
Cloudamize\\
Philadelphia, PA 19103\\
stephan@cloudamize.com}}


%Telephone: (800) 555--1212, Fax: (888) 555--1212}


\IEEEoverridecommandlockouts
\IEEEpubid{\makebox[\columnwidth]{978-1-5090-4026-1/17/\$31.00~
        \copyright2017
        IEEE \hfill} \hspace{\columnsep}\makebox[\columnwidth]{ }}

\maketitle

\begin{abstract}
In the cloud environment, data centers are efficiently manipulated by cloud service providers (CSPs) in terms of energy consumption. Consequently, migrating workloads to clouds can result in lower energy consumption. This paper demonstrates that the Lift-and-Shift migration with optimal selections of cloud instances can provide significant energy savings, and explains how much and where the energy savings are obtained from. Additionally, the analysis on the variation of energy consumption is given when Auto-Scaling is deployed showing that further energy savings are expected even without refactoring applications. These findings
contract the popular belief that one needs to modify applications
in order to achieve the benefits of the cloud. All the conclusions and analyses are based on the real data collected by Cloudamize Inc. from May 2016 to August 2016 over 40,000 machines across approximately 300 data centers.
\end{abstract}

\begin{IEEEkeywords}
Cloud computing, data migration, Lift-and-Shift, Auto-Scaling.
\end{IEEEkeywords}

% \fi
\IEEEpeerreviewmaketitle

\section{Introduction}
The cloud computing paradigms, e.g. IaaS, offer a wide range of benefits on running workloads over traditional data centers (TDCs), including easy deployment, redundancy, and access to globally deployed infrastructures. Modern CSPs have developed and deployed provenly efficient computing systems \cite{awsCustomHardware}. The data centers manipulated by CSPs are running far more efficiently than traditional ones. One key reason is that the energy consumption can be significantly lowered down by migrating workloads from TDCs to clouds. In this paper, we demonstrate that simple Lift-and-Shift strategies with optimal selections of cloud instances can save energy significantly even without refactoring applications. More specifically, the energy usage can be reduced by a factor of 4.5 to 7.8, and even 3 to 5 times more potentially when Auto-Scaling (with refactoring) could be deployed. The data in use is all real and collected by Cloudamize Inc.\cite{cloudamize}, which makes this study, to the best of our knowledge, be the first large one that quantifies the energy savings in the real cases.

\begin{figure}[h]
    \centering
    \includegraphics[width=3in]{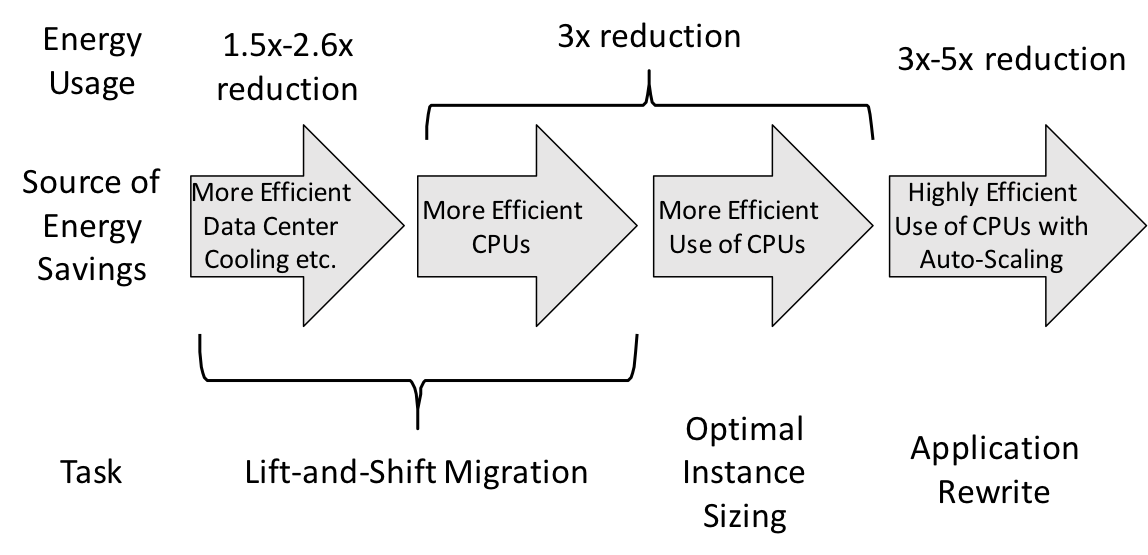}
    \caption[Energy Reduction from Migrating to the Cloud and \newline\ Optimizing Usage
    of the Cloud Services]{Energy savings pipeline}
    \label{fig:fig1_eng_reduct}
\end{figure}

Figure \ref{fig:fig1_eng_reduct} provides an overview of the energy savings pipeline, the sources and the tasks required to achieve the savings. The application migration is divided into three parts, lift-and-shift, optimal instance sizing, and application rewrite. 

The organization of this paper is shown as follows. In Section \ref{sec:LiftAndShift}, we quantify the energy savings by using the lift-and-shift strategies. In Section \ref{sec:optSize}, we explain how the optimal instance sizing should be done. Section \ref{sec:autoScaling} shows the analysis on how Auto-Scaling could affect the energy consumption. And finally, Section \ref{sec:conclusion} summarizes our remarked conclusions.

\section{Related Work}
\label{sec:related}
Cost saving is one of the core motivations of migrating data from TDCs to clouds. Armbrust et al. \cite{armbrust2010view} propose that the elasticity and transference of the risks of over-provisioning and under-provisioning are the most important economic benefits of cloud computing. Walker et al. \cite{walker2009real} make a comparison of CPU hours between clouds and traditional infrastructure, and propose an optimal expected CPU utilization at 40\%. Another case study \cite{khajeh2010cloud} also shows that the migration of an IT system from an in-house data center to the cloud reduced 37\% cost in 5 years. These studies are all based on economic or business perspectives to help make decisions, yet very few studies investigate into quantifying energy savings of cloud migration.

Lawrence Berkeley National Laboratory provides an open-access energy-efficiency model, Cleer \cite{masanet2013energy}, to estimate the energy savings when moving current business software to clouds. The study concludes that the potential primary energy savings can hit up to 23 billion kWh/yr (i.e. 87\% in reduction); however, this model merely provides the estimation in three specific scenarios without general considerations on system specifications. Additionally, this study only focuses on total energy consumption yet leaving the efficiency missed. 

\section{Summary of the data set}
\label{sec:dataSet}
The data used in this study is collected by Cloudamize Inc.\cite{cloudamize} from May 2016 to August 2016. This data set includes a wide range of performance metrics from more than 40,000 machines (both virtual and physical) in approximately 300 data centers. The data collection lasted 14 days in minimum and 21 days in average. For each machine, the CPU utilization was recorded every 20 or 30 seconds depending on the system. The underlying hardware information was also collected, e.g. CPU details, so was the network usage information including the source and destination IP addresses of packets.

Cloudamize is a commercial cloud computing company that provides system migration and management services in clouds. The data all comes from the customers or partners of Cloudamize who used TDCs and have been considering to migrate to clouds. Cloudamize finds that the involved cases are all originated from business concerns such as reducing costs, improving application development cycle, improving agility, decoupling from maintaining computing infrastructure and focusing on core business. The applications are not designed dedicated to the cloud environments and varying from areas to programming. Thus it should be considered as valid to conduct our analysis.

\section{Lift-and-Shift and Energy Consumption}
\label{sec:LiftAndShift}
Since the lift-and-shift migration is in use and the computational capabilities in the cloud end will be the same as the on-premise hardware, the difference of energy consumption between these two environments is actually a matter of how efficiently the resources are utilized. In this section, we first quantify and evaluate the computational efficiency, then we use the efficiency to estimate the change in the energy usage.

Over the past several decades, the techniques on the energy saving have gained a huge progress, and new techniques are emerging every day. In this situation, hardware updates have become increasingly important to both TDCs and CSPs, and usually CSPs can keep higher update rates than TDCs can do \cite{whitney2014data}. Hence, it could be believed that migrating workloads to clouds can save energy significantly meanwhile keeping up the computational performance.

We define the \textit{computational efficiency} (CE) of a CPU to be the ratio of the SPECCPU2006. SPECCPU2006 benchmark is widely used and has been evaluated on nearly every CPU released by Intel. benchmark \cite{SPECCPU2006} and the TDP of the CPU. Let $CE_{datacenter}$ be the CE for the on-premise machine and $CE_{cloud}$ be the CE for the cloud hardware. The ratio $CE_{datacenter}/CE_{cloud}$ estimates the fraction of energy consumed after the workloads being migrated from TDCs to the cloud. 

\begin{figure}[h]
    \centering
    %\vspace*{-0.2in}
    \includegraphics[width=2.5in]{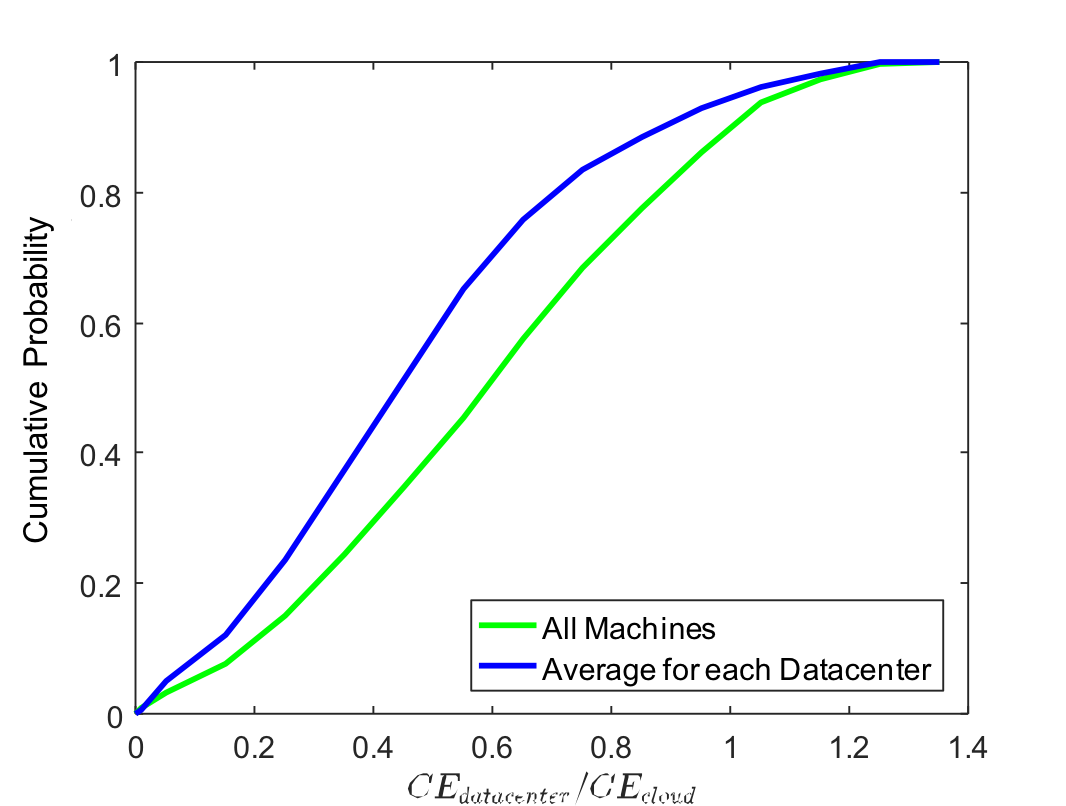}
    \caption{Fraction of Energy Consumed After Migration}
    \label{fig:CERatio}
\end{figure}

In Figure \ref{fig:CERatio}, the mean fraction of energy consumed is 0.64 over all machines and 0.51 for TDCs. The figure shows that not all workloads would experience a reduction of the energy usage. Around 10\% of all machines would use more energy after switching to the CPUs used by CSPs, (i.e. around 5\% of TDCs would see an increase of the average energy usage). The reason is that some CPUs used in TDCs are more updated than those in the cloud. However, in Section \ref{sec:optSize}, we will see that if these machines are sized more precisely, the energy usage will decrease for nearly all machines. The figure also shows that a half of the machines would use around 40\% less energy if the workloads are simply lift-and-shifted, and 50\% of the TDCs would experience an average 50\% reduction in energy usage. One possible explanation of the difference between the average on data centers and the average on individual machines is that the statistical result is biased due to a number of small inefficient ones. 

\begin{figure}[h]
    \centering
    %\vspace*{-0.3in}
    \includegraphics[width=2.5in]{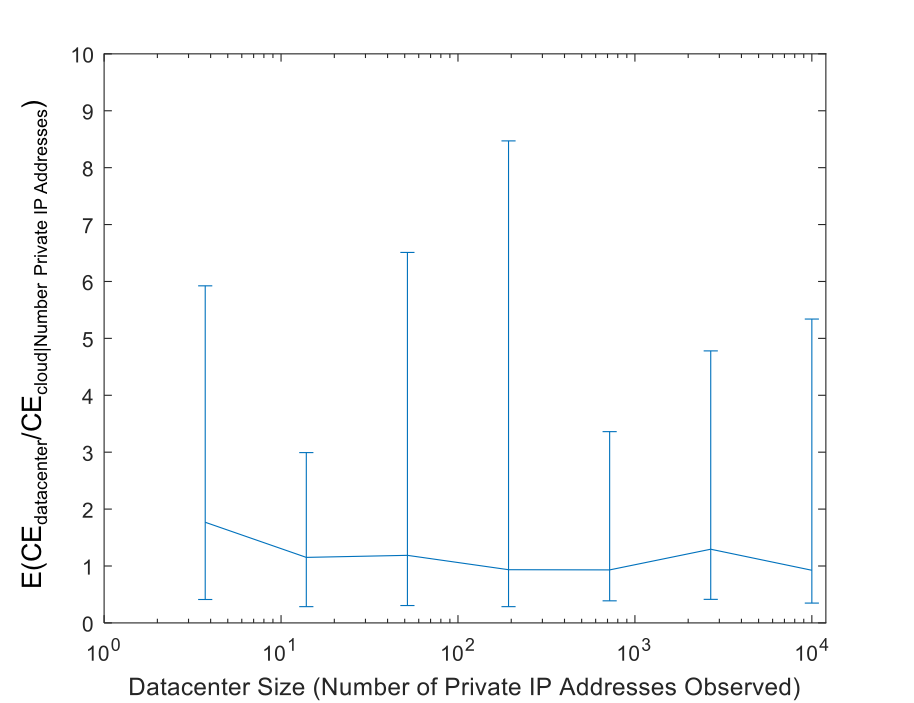}
    \caption{Average Fractions of Energy Usage on Different TDC Sizes}
    \label{fig:CeDsSize}
\end{figure}

Figure \ref{fig:CeDsSize} explores this hypothesis. Data centers are grouped into seven bins by their private IP addresses. For each group, the mean value, maximum and minimum value of $CE_{datacenter}/CE_{cloud}$ are computed. This figure indicates that there is no significant dependence between $CE_{datacenter}/CE_{cloud}$ and the size of the data center, (i.e. the size of data center would not influent the energy consumption significantly).

%-------------------Figure-------------------
%--------------------------------------------
%-------------------Figure-------------------
%--------------------------------------------

\section{Optimal Instance Sizing}
\label{sec:optSize}

Several studies have indicated that CPU over-provisioning is an important factor of energy waste in data centers \cite{whitney2014data}\cite{armbrust2010view}\cite{judge2008reducing}. However, the conclusion is not based on direct measures of CPU utilization. The cloud offers several methods that can greatly reduce energy waste (and cost) caused by the over-provisioning. This section studies the case that the instance type is rarely changed, and examines the potential energy savings by choosing the optimal sized instances. 

\subsection{CPU Utilization and Energy Consumption}
\label{subsec:CErelation}

CSPs provide a large number of instance types, each with different computational capabilities. For example, Azure currently offers over 70 types; Google Cloud also offers a wide range of instance types, as well as customized instance types, with which the user can specify the number of CPU cores, the amount of memory and easily resize the type. The instance types in cloud make the resizing much simpler than it is in TDCs. 

Even though applying the virtualization in TDCs allows to resize the machines, it does not solve the basic problem that the computational capabilities must be purchased and installed before usage. Data center resizing is usually a complicated process that might require approval from several levels of management, and updating other facilities such as power and cooling. Thus, usually additional capacities will be purchased so that to guarantee the data center resizing to be infrequent. As a result, the corresponding system design may lead to a low CPU utilization. The target CPU utilization is not based on performance objectives, but a trade-off solution on performance goals, anticipated growth, and the desired resizing cycle.

\begin{figure}[h]
    \centering
    \includegraphics[width=3.0in]{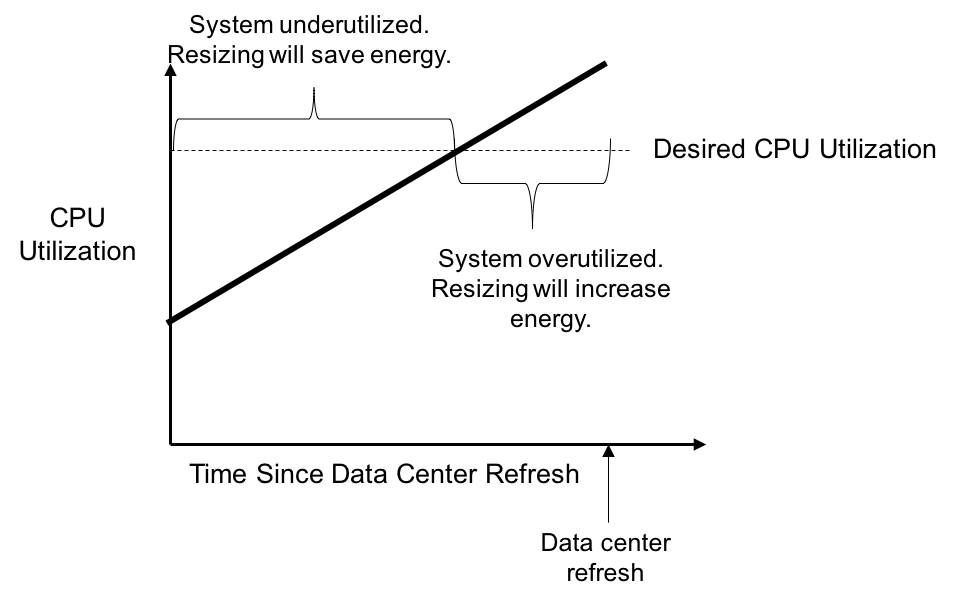}
    \caption{Periodic Data Center Refresh and Potential Impact on Energy Savings when Migrating to the Cloud}
    \label{fig:perodicRefresh}
\end{figure}

The result is that systems in TDCs can experience a cyclic-mannered CPU utilization. Initially, after the data center is sized, the CPU utilization is low and the hardware is underutilized. After a certain amount of time, the hardware utilization becomes more efficient, and eventually reaches a level of high degree before another data center resizing is performed and then the cycle starts over again. Figure \ref{fig:perodicRefresh} illustrates how the system utilization might vary between the data center updates. There are also a wide range of other factors that impact the utilization and performance. The performance of applications is only one of many such factors. Hence, the utilization in data centers could be too low or too high, either of which is far from the optimal solution. 
%-------------------Figure4-------------------
%-------------------Figure-------------------

\begin{figure}[h]
    \centering
    \includegraphics[width=2.5in]{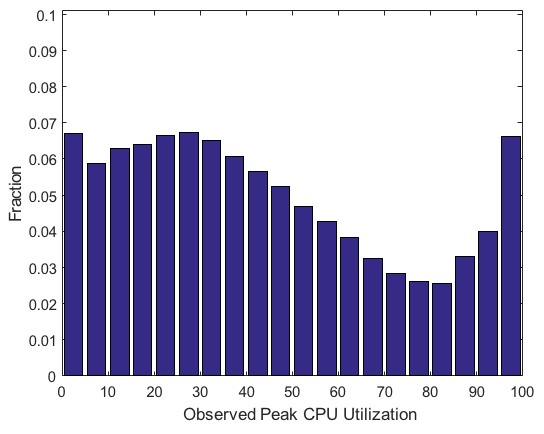}
    \caption[Cumulative Probability Distribution of Observed CPU Utilization]{Cumulative Probability Distribution of CPU Utilization}
    \label{fig:cpuCDF}
\end{figure}

\begin{figure}[h]
    \centering
    \includegraphics[width=2.5in]{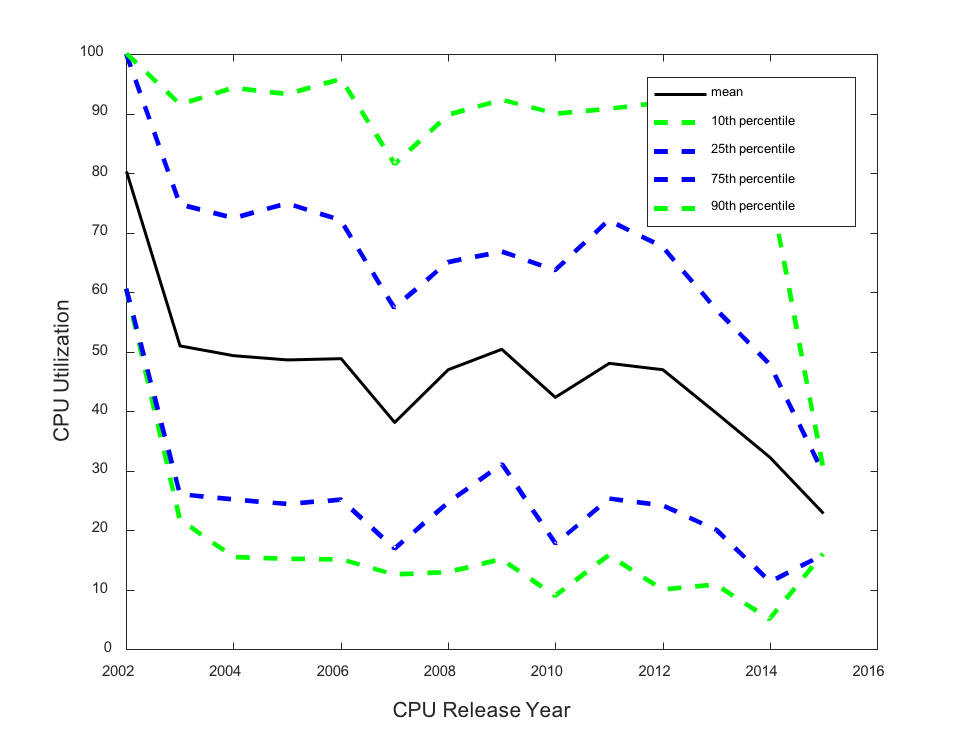}
    \caption[CPU Utilization vs CPU Release Date]{CPU Utilization vs CPU Release Date}
    \label{fig:cpuVSReleaseDate}
\end{figure}

Figure \ref{fig:cpuCDF} shows the \textit{peak CPU utilizations} observed. The CPU utilization is measured every 20 or 30 seconds (depending on the system settings). These high-frequency measurements are smoothed with a 5-minute smoothing window. The maximum value of this utilization is collected for each day. The \textit{peak CPU utilization} is the 95th percentile for these daily maximum values. This method is selected so that single busy periods have little impact on the estimation of whether the system is over or underutilized. The figure also shows that the \textit{peak CPU utilization} is frequently and fairly low. On the other hand, a non-negligible fraction of machines have high peak CPU utilization. Figure \ref{fig:cpuVSReleaseDate} shows the mean CPU utilization along with the 10th, 25th, 75th, and 90th percentiles of the CPU utilization as a function of the release date of the CPU. This figure indicates that the \textit{peak CPU utilization} of CPUs released since 2013 is typically lower than the peak utilization of CPUs released before 2013. This observation agrees with the data center refresh model described above, in which machines in data centers with updated hardware are expected to have lower utilizations than machines with out-of-date hardware.

It is easy to see from the figures that many machines are over-provisioned, and thus energy can be saved by moving the workloads to the machines with less computational capabilities. During migration, the machines with high CPU utilization would be moved to the machines with more computational capabilities, which would increase the energy usage compared to a lift-and-shift migration that seeks to keep CPU utilization unchanged. In order to quantify the change in energy usage from resizing the machines, we need to 1) select a target peak CPU utilization, and 2) estimate how energy usage changes with CPU utilization. Determining the optimal CPU utilization is a complicated issue that is standing outside the scope of this paper. Moreover, system designers select the target peak CPU utilization based on various "rules-of-thumb." Therefore, in the following analysis, we consider the target CPU utilizations ranging from 50\% to 90\%. Additionally, estimating how energy usage would be varying with CPU utilization is discussed in the next section.
%-------------------Figure 5-------------------
%-------------------Figure-------------------

%-------------------Figure 6-------------------
%-------------------Figure--------------------

\subsection{Energy Consumption as a Function of CPU Utilization}
\label{subsec:CEmodel}

\begin{figure}[h]
    \centering
    \includegraphics[width=2.5in]{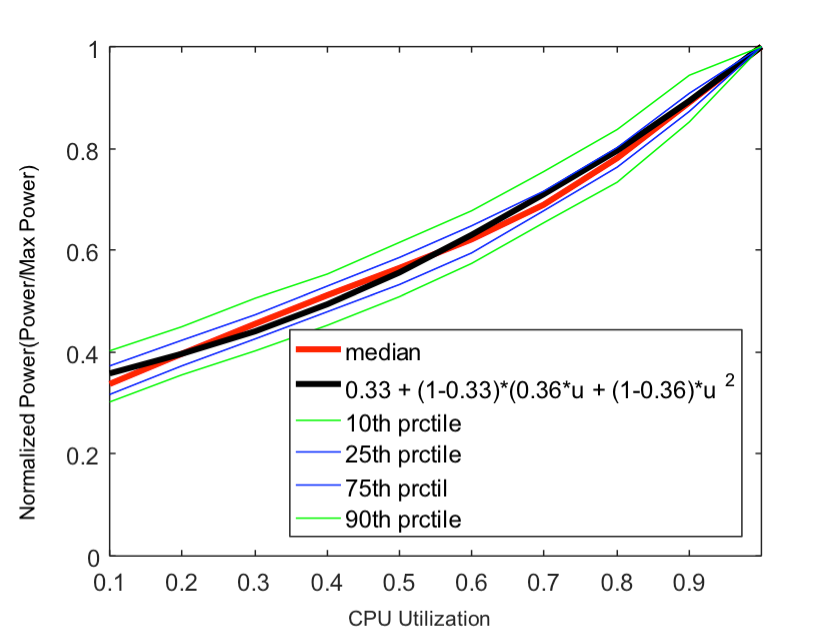}
    \caption{System Energy Usage vs CPU Utilization from the SPECpower\_ ssj2008 for CPUs Released after 2012}
    \label{fig:specEnergy2}
\end{figure}

To understand the relationship between the energy usage and the CPU utilization, we introduce SPECpower\_ssj2008 benchmark\cite{SPECpower2008}. Among the submitted results, 488 samples of Intel Xeon have been submitted, covering 86 different Intel Xeon CPUs released from 2005 to 2016. The data reveals that the energy consumption on contemporary processors are not proportional to their utilization. Particularly, for example, their idle power consumption is surprisingly high compared to the full-load power consumption. Figure \ref{fig:specEnergy2} shows the relationship between the energy usage and the CPU utilization for CPUs released after 2012. This figure also includes the graph of the function
$
E\left( u\right) =0.33+\left( 1-0.33\right) \left( 0.36u+\left(
1-0.36\right) u^{2}\right) .  
\label{eq:EvsCpu}
$

Different from the model developed in \cite{powervsCPUModel}, our model focuses and is only applicable on recently released CPUs. The CPUs used by current CSPs such as Azure, GCP, and AWS are all recently released; therefore the model $E\left( u\right)$ can be used to estimate the relative energy usage as a function of CPU utilization on systems that have been migrated to the cloud.

%-------------------Figure 7-------------------
%-------------------Figure-------------------
\subsection{Energy Consumption with Statically Sized Instances}
\label{subsec:statical}
With the model $E\left( u\right)$, we can estimate the change of the energy usage resulted from resizing the machine, and focus on statically sized machines. We select a single machine size, and examine the energy usage where the computational resources allocated to a workload are allowed to vary dynamically. Using a single machine size is of considerably less administrative effort than dynamically sizing machines. However, statically sizing machines means that the machine size is rarely changed, though not never changed. We assume that the machine is not resized to accommodate the workload in the collected data. That is, we will select a single instance size for each workload, where the instance size must be suitable for the entire data set for that workload. This is a reasonable assumption because the typical data set is only for a few weeks and does not include any data set that spans longer than two months.

Let $u\left( t\right) $ be the CPU utilization observed at time $t$. Then, the normalized energy usage is $\int E\left( u\left( t\right) \right) dt$; this is the normalized energy usage when the system is migrated (i.e. via a lift-and-shift migration) to an instance that has the exactly same computational capabilities as the hardware used in the on-premise data center. We assume a simple scaling model, where the computation capabilities vary linearly with the amount of computation resources applied. For example, if a system is allocated with two cores, then allocating one core will exactly double the CPU utilization (up to a maximum of 100\%). We also assume that fractional resources (e.g. 2.4 cores) can be allocated. Moreover, we assume that the energy usage scales linearly with the number of resources allocated for a fixed CPU utilization. This means that if a system is 50\% utilized and if we double the computational resources allocated to the system while the load is doubled, the energy usage is also doubled.

With these assumptions, we have the following model of energy usage when the computational capabilities are increased by a factor of $c$, $E\left( u\left( t\right)/c\right) \times c$. The CPU utilization decreases by a factor of $c$, which leads to a change in the energy usage. However, the number of machines also increases by a factor of $c$, which increases the energy usage by a factor of $c$. Note that if the energy usage is linear on the CPU utilization, $E\left( u\left( t\right) /c\right) \times c=E\left( u\left( t\right) \right) $, and correctly sizing cloud instances would not impact the energy usage. However, despite efforts, the energy usage is not a linear function of the CPU utilization\cite{ganapathy2008defining}.

We denote $u_{p}$ as the observed \textit{peak CPU utilization} and $u_{T}$ as the target \textit{peak CPU utilization}. Since we assume a simple scaling model, we can achieve $u_{T}$ by adjusting the computational resources allocated by a factor of $c=u_{p}/u_{T}$. In this case, the fraction of energy usage after optimally resizing the cloud instance is
$
\nicefrac{\int E\left( u\left( t\right) /\left( u_{p}/u_{T}\right) \right)
\times \left( u_{p}/u_{T}\right) dt}{\int E\left( u\left( t\right) \right) dt
}.
$

\begin{figure}[h]
    \centering
    \includegraphics[width=2.5in]{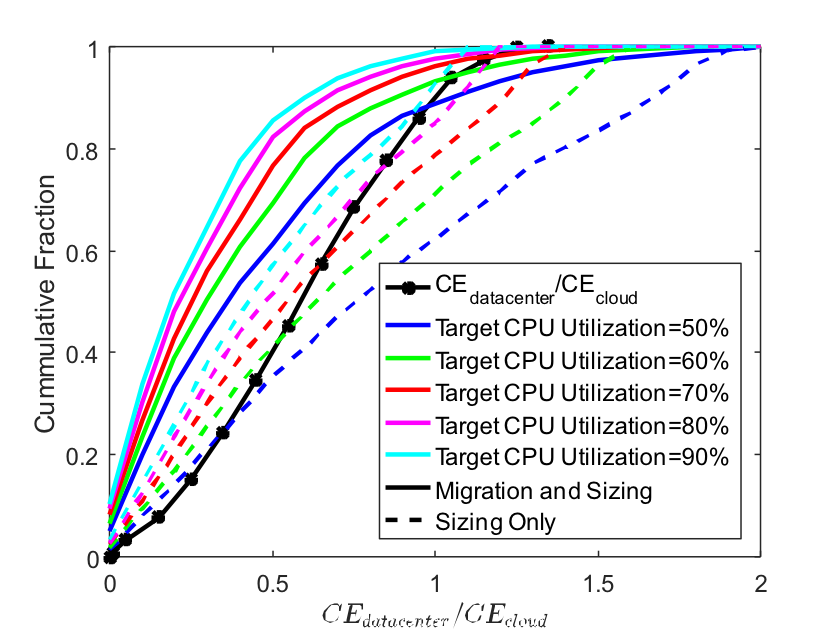}
    \caption{Fraction of energy used after migrating the workload to the cloud
        via a lift-and-shift migration and then resizing the optimally sized instance}
    \label{fig:resizeAndMig}
\end{figure}

Figure \ref{fig:resizeAndMig} compares the fraction of energy by performing a lift-and-shift alone, optimal resizing alone, or the combination of these two strategies. The figure shows that resizing the instance type does not reduce the energy in all cases. This is expected since Section \ref{subsec:CErelation} shows that energy usage is quite large on some machines. For larger target CPU utilization, a larger fraction of system experience a reduction in energy usage as a result of resizing. Recall in Section \ref{sec:LiftAndShift}, the energy usage does not decrease for all workloads when utilizing a lift-and-shift migration if the hardwares are newer in data centers. But these machines tend to be underutilized. Therefore, a combination of the lift-and-shift and the instance resizing results in significant energy savings, and greatly exceeds the energy saving of either of the lift-and-shift migration and the resizing. Table \ref{tab: MeanEngMig} shows the mean fraction of energy used after this type of migration.

%-------------------Figure 8-------------------
%-------------------Figure-------------------
%-------------------Table-------------------
\begin{table}[!ht]
\renewcommand{\arraystretch}{1.3}
\centering
\caption{Mean Fraction of Energy Used after Migration and Resizing}
\label{tab: MeanEngMig}
\begin{tabular}{|c|c|}\hline
Target Peak CPU Utilization & Mean Fraction of Energy Used \\\hline
50\% & 0.51 \\ \hline
60\% & 0.43 \\ \hline
70\% & 0.37 \\ \hline
80\% & 0.33 \\ \hline
90\% & 0.30 \\ \hline
\end{tabular}
\end{table}
%-------------------Table-------------------

The energy saving identified is significant. For reasonable values of target peak CPU utilization of 70\%-80\% (the other values are included for references), the lift-and-shift with an optimal sized instance reduces energy usage by nearly a factor of 3. Moreover, there exists a wide range of tools that nearly automate this type of migration \cite{cloudEndure} \cite{riverMeadowWebSite} \cite{MsftAsrOnline} \cite{cloudamize}.

\section{Auto-Scaling and Energy Consumption}
\label{sec:autoScaling}
The IaaS paradigm has many useful characteristics. One is that users are charged based on the duration that machines are running. As a result, there are significant financial advantages to apply the auto-scaling. This section examines the potential energy savings where computational abilities are frequently adjusted through auto-scaling.

Here is a simple idealized model for energy usage when the auto-scaling is employed. Let $c\left( t\right) $ be the resources allocated to the workload at time $t$. Then, let \thinspace $u\left( t\right) $ be the observed CPU utilization of the workload in the data center. The CPU utilization in the cloud at time $t$ would be $u\left( t\right) /c\left( t\right) $, while the energy usage would be $E\left( u\left( t\right)/c\left( t\right) \right) \times c\left( t\right) $. Let $u_{T}$ be the target CPU utilization. Then, we set $c$ so that $u_{T}=u\left( t\right) /c\left( t\right) $. Therefore, the energy used is $E\left( u_{T}\right) \times u\left( t\right) /u_{T}$. The fraction of energy used by the auto-scaling as compared to the energy usage after a lift-and-shift migration is
$
\nicefrac{\int E\left( u_{T}\right) \times u\left( t\right) /u_{T} \ dt}{\int
E\left( u\left( t\right) \right) \ dt}.
$
%-------------------Figure 9-------------------
\begin{figure}[h]
\centering
\includegraphics[width=2.5in]{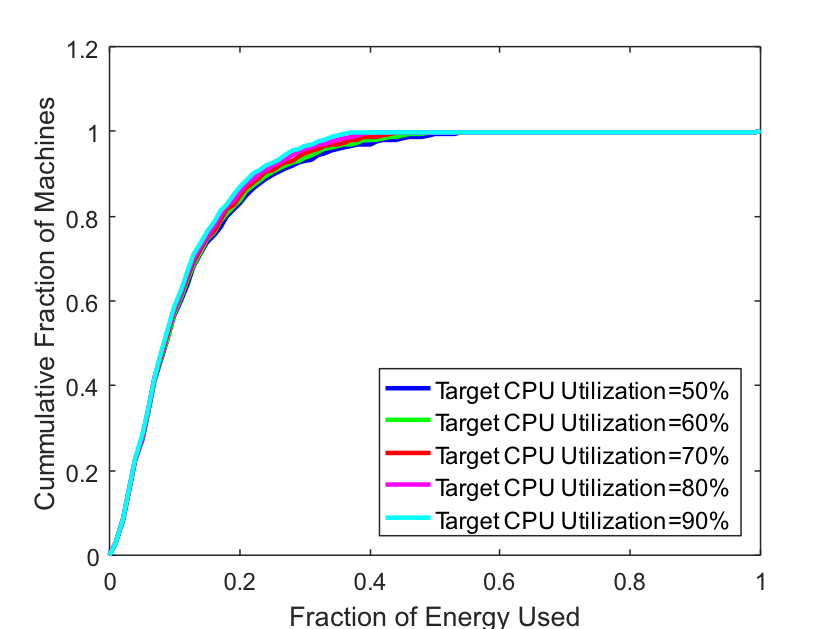}
\caption{Fraction of Energy Used After Employing Idealized Auto-Scaling}
\label{fig:idealAS}
\end{figure}
%-------------------Figure-------------------
%-------------------Figure 10-------------------
\begin{figure}[h]
\centering
\includegraphics[width=2.5in]{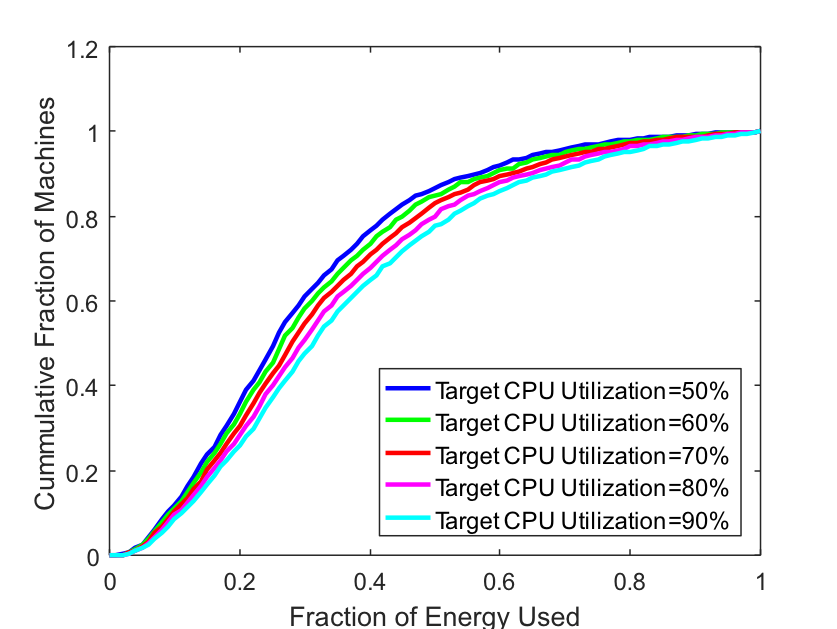}
\caption{Fraction of Energy Used After Employing Idealized Auto-Scaling where the Instance is Resized Once an Hour}
\label{fig:oneHrAS}
\end{figure}
%-------------------Figure-------------------

Figure \ref{fig:idealAS} shows the cumulative distribution of the fraction of energy used after the auto-scaling is applied, as compared to the energy used after migrating to the cloud and optimally sizing the instance type. Clearly, significant energy savings are possible. However, there are significant drawbacks of this model. First, not all workloads are suitable for auto-scaling. Even suitable workloads might require a significant rewrite. Second, the model that the allocated computational resources, $c\left(t\right) =u\left( t\right) /u_{T}$, is highly idealized. Specifically, it assumes that the computational resources are changed instantaneously and continuously. This type of energy usage might be reasonable for a PaaS system such as AWS’s Lambda service, yet it perhaps too idealized for most workloads and IaaS.
% \begin{figure}[h]
%     \centering
%     \includegraphics[width=3.6in]{fig_9_10.png}
%     \caption{Fraction of Energy Used After Employing Idealized Auto-Scaling where the Instance is Resized Once an Hour}
%     \label{fig:AS}
% \end{figure}

Figure \ref{fig:oneHrAS} shows the energy savings in a slightly more realistic scenario. Here we assume that 
$
c\left( t\right) =(\nicefrac{1}{u_{T}})\max_{s\in hour\left( t\right) }u\left(
s\right) , 
$
where $hour\left( t\right) $ is the hour of time $t$. Clearly, this model suffers from the drawback where it assumes the CPU utilization can be accurately predicted. Methods to predict CPU utilization and the impact of inaccuracy are out of scope of this study. However, there are likely to be significant energy savings by refactoring applications to take advantage of auto-scaling. For example, the mean fraction of the energy used is around 0.3, implying a reduction of energy usage by a factor of 3.

\section{Conclusion}
\label{sec:conclusion}

This paper presents results regarding energy savings that will be achieved by migrating workloads to the cloud. The data indicates that such migrations will reduce energy usage by a factor of between 4.5 and 7.8. Relatively little energy savings come from the efficiency of CSPs data centers in terms of efficient cooling and lighting systems. The dominant amount of savings is benefited from the lift-and-shift and the optimal instance sizing. These sources of energy savings are more beneficial significantly than moving from less efficient data centers to efficient ones owned by CSPs. It is critical to note that the migration and the resizing can be preformed nearly automatically \cite{cloudEndure}\cite{riverMeadowWebSite}\cite{MsftAsrOnline}\cite{cloudamize}. 

The sources of the energy are also important concerns. For example, CSPs often utilize solar and wind farms to offset their energy usage. As a result, migrating workloads to the cloud could reduce carbon emissions beyond what would be achieved by only reducing energy usage. Further study is required to clarify and quantify the influence on carbon emission caused by the cloud migration.

\bibliographystyle{./IEEEtran}
\bibliography{./IEEEabrv,./biblist}

\begin{thebibliography}{10}
\providecommand{\url}[1]{#1}
\csname url@samestyle\endcsname
\providecommand{\newblock}{\relax}
\providecommand{\bibinfo}[2]{#2}
\providecommand{\BIBentrySTDinterwordspacing}{\spaceskip=0pt\relax}
\providecommand{\BIBentryALTinterwordstretchfactor}{4}
\providecommand{\BIBentryALTinterwordspacing}{\spaceskip=\fontdimen2\font plus
\BIBentryALTinterwordstretchfactor\fontdimen3\font minus
  \fontdimen4\font\relax}
\providecommand{\BIBforeignlanguage}[2]{{%
\expandafter\ifx\csname l@#1\endcsname\relax
\typeout{** WARNING: IEEEtran.bst: No hyphenation pattern has been}%
\typeout{** loaded for the language `#1'. Using the pattern for}%
\typeout{** the default language instead.}%
\else
\language=\csname l@#1\endcsname
\fi
#2}}
\providecommand{\BIBdecl}{\relax}
\BIBdecl

\bibitem{awsCustomHardware}
Amazon{W}eb{S}ervices, ``A{WS} re:{I}nvent 2015 | ({SPOT}301) {AWS}
  {I}nnovation at {S}cale,'' \url{https://www.youtube.com/watch?v=uIjh79NeSVQ},
  2015.

\bibitem{cloudamize}
Cloudamize, \url{http://www.cloudamize.com/}.

\bibitem{armbrust2010view}
M.~Armbrust, A.~Fox, R.~Griffith, and et~al, ``A view of cloud computing,''
  \emph{Communications of the ACM}, vol.~53, no.~4, pp. 50--58, 2010.

\bibitem{walker2009real}
E.~Walker, ``The real cost of a cpu hour,'' \emph{Computer}, vol.~42, no.~4,
  2009.

\bibitem{khajeh2010cloud}
A.~Khajeh-Hosseini, D.~Greenwood, and I.~Sommerville, ``Cloud migration: A case
  study of migrating an enterprise it system to iaas,'' in \emph{2010 IEEE 3rd
  International Conference on cloud computing}.\hskip 1em plus 0.5em minus
  0.4em\relax IEEE, 2010, pp. 450--457.

\bibitem{masanet2013energy}
E.~Masanet, A.~Shehabi, L.~Ramakrishnan, and et~al, ``The energy efficiency
  potential of cloud-based software: A us case study,'' \emph{The Energy
  Efficiency Potential of Cloud--Based Software: A US Case Study}, 2013.

\bibitem{whitney2014data}
J.~Whitney and P.~Delforge, ``Data center efficiency assessment,'' \emph{Issue
  Paper, Aug}, 2014.

\bibitem{SPECCPU2006}
{SPEC}, ``S{PEC} {CPU}2006,'' \url{https://www.spec.org/cpu2006/}.

\bibitem{judge2008reducing}
J.~Judge, J.~Pouchet, A.~Ekbote, and et~al, ``Reducing data center energy
  consumption,'' \emph{ASHRAE Journal}, vol.~50, no.~11, p.~14, 2008.

\bibitem{SPECpower2008}
{SPEC}, ``S{PEC}power\_ssj2008,'' \url{https://www.spec.org/power\_ssj2008/}.

\bibitem{powervsCPUModel}
X.~B. Fan, W.~D. Weber, and L.~A. Barroso, ``Power provisioning for a
  warehouse-sized computer,'' in \emph{ACM SIGARCH Computer Architecture News},
  vol.~35, no.~2.\hskip 1em plus 0.5em minus 0.4em\relax ACM, 2007, pp. 13--23.

\bibitem{ganapathy2008defining}
D.~Ganapathy and E.~J. Warner, ``Defining thermal design power based on
  real-world usage models,'' in \emph{Thermal and Thermomechanical Phenomena in
  Electronic Systems, 2008. ITHERM 2008. 11th Intersociety Conference
  on}.\hskip 1em plus 0.5em minus 0.4em\relax IEEE, 2008, pp. 1242--1246.

\bibitem{cloudEndure}
Cloud{E}ndure, \url{https://www.cloudendure.com/}.

\bibitem{riverMeadowWebSite}
River{M}eadow, \url{http://www.rivermeadow.com/}.

\bibitem{MsftAsrOnline}
Microsoft, ``Microsoft {A}zure,''
  \url{https://docs.microsoft.com/en-us/azure/}.

\end{thebibliography}

\end{document}